# SOUND VELOCITY IN LIQUID AND GLASSY SELENIUM

W.B. Payne[‡], J.K. Olson[‡], A. Allen[‡], V.F. Kozhevnikov[‡†*] and P.C. Taylor[‡]

[‡]University of Utah, Salt Lake City, UT 84112, USA
[†]Katholike Universiteit Leuven, B-3001 Leuven, Belgium.

The speed of longitudinal sound waves at 7 and 22 MHz has been measured in liquid, supecooled, and amorphous selenium, including the region around the glass transition temperature, $T_g$, near 35°C. In amorphous selenium the speed of shear waves at 7 MHz was also measured. The experiments were performed with high purity Se (99.9999%) hermetically sealed in an evacuated quartz ampoule. Four temperature regions with strongly different relaxation times can be distinguished between room temperature and the melting point: (1) a glassy state below $T_g$, which is stable on the time scale of the experiments, (2) a glassy state above $T_g$, which is metastable on the time scale of the experiments, (3) a region where homogeneous crystal nucleation occurs, and (4) a supercooled liquid, which is stable on the time scale of the experiments. Each region is marked by a change in the slope of the temperature dependence of the sound velocity. Near the glass transition temperature the velocities of longitudinal and transverse sound exhibit hysteresis with a step-like drop on heating and a more continuous rise on cooling. The step-like anomaly in sound velocity may be a general property of the glass transition.

## 1. INTRODUCTION

During the last decade significant progress has occurred in understanding the physical properties of supercooled liquids, amorphous solids, and the transition between these two states [1-4]. However, the complexity of these materials has insured that many challenging questions remain, such as the question of possible phase transitions in the supercooled state [5, 6].

Pure selenium is the simplest glass-forming substance, and therefore most attractive for modeling. Like sulfur, selenium consists of a wide variety of allotropes of cyclic molecules and chain polymers [7, 8]. But in contrast to sulfur, in which the polymers are essential mainly in the liquid state [9-11], the polymers in selenium are significant in both the solid and liquid phases [12-14]. Selenium is used for many practical applications, as a major glass-forming component in many chalcogenide glasses and as an important material for opto-electronics [15]. Because of its practical importance, many experimental, theoretical, and computational studies of the physical properties of selenium have been published (see, for example, Refs 16, 17, and references contained therein). Amorphous selenium undergoes a glass transition at about 35 °C, where the heat capacity exhibits a step-like anomaly [18], the viscosity is on the order of $10^{13}$ poise [19-21], and the temperature derivative of the mean-square atomic displacement exhibits a drastic change [22, 23].

[*]Corresponding author. E-mail: Vladimir.Kozhevnikov@fys.kuleuven.ac.be

Sound velocity, because of its direct association with the elastic properties, can provide further insight into the supercooled liquid and the glass transition. In some glassy systems, the sound velocity exhibits a step-like drop near the glass transition temperature [4], but it is not clear if this property is universal. In particular, molecular-dynamic simulations [24] suggest that the glass transition may be similar to a polymerization transition. If so, near the glass transition one expects to find a change in the slope of the temperature dependence of the sound velocity, as occurs in sulfur at the polymerization transition [9, 11].

The primary goal of this work is to determine for selenium what specific features exist near the glass transition temperature in the speeds of longitudinal and transverse sound. Another goal is to construct a "phase diagram" of supecooled selenium based on measurements of sound velocity. This phase diagram should contain the limiting temperatures for the regions of relative stability of the superheated glass and the supercooled liquid. It is difficult to quantify these limits because they depend on the heating and cooling rates. Neutron scattering experiments [23] have established the upper limit for the superheated glass and the lower limit for the supercooled liquid to be 360 and 440 K, respectively. In between these two limits the system will crystallize on a specific time scale, which for our measurements we take to be less than about 10 s. The present experiments examine the temperature dependence of the sound velocity near these two limits to look for any characteristic features.

Data on sound velocity in selenium have been reported for the liquid state at temperatures up to 550 °C [25] and for the glassy state at 25 °C [26]. Sound velocity data are also available for liquid Se-Te mixtures at temperatures up to about 830 °C [27].

2. EXPERIMENTAL

The cell used in this work has the same design as the cell we previously employed for sulfur [9]. This cell consists of a quartz tube and two quartz buffer rods, which are hermetically sealed at the ends of the tube. High purity selenium (99.999%, Alfa Aesar) was sublimed into the cell, after which the cell was sealed off under vacuum. The sample was 12 mm in diameter and about 3 mm in length. The sample length was measured with an uncertainty of 2 μm. Details are available elsewhere [9].

The cell was installed in a silicon oil bath connected to a thermostat (6330, Hard Scientific, Inc) supplied with a homemade pump. An electrical heater made of nickel-chromium wire was wound about the bath. The power of the heater was controlled with a phase angle power controller PC 120/240-PA and Smart-3 temperature controller (both from Temp, Inc). The entire bath was covered with alumina wool for thermal isolation. The sample temperature was measured using three cooper/constantan thermocouples fixed on the cell body near the sample space. The temperature gradient over the sample was less than 0.1 °C.

Single crystal lithium niobate and quartz piezoelectric transducers were used for measurements of longitudinal and shear waves, respectively. The resonant frequency of the transducers was 7 MHz. The transducers for the measurements of longitudinal sound were attached to the outer ends of the buffer rods using unpolymerized epoxy resin. Polymerized resin was used to bond the shear transducers. The measurements were performed using a pulsed phase sensitive technique described elsewhere [28]. Changes were measured in the propagation time of the sound through the sample with respect to a reference point. These changes had an uncertainty of $2 \times 10^{-10}$ s. The reference point was chosen at 25 °C, where the values for velocity of longitudinal and transverse sound from Ref. 26 were assumed. The accuracy of the absolute data for the sound velocity is mostly determined by the uncertainty in the literature values [26] we employed at the reference point. We estimate this accuracy to be about 0.2-0.3 %.

For the experiments with supercooled liquid selenium, the sample was melted and heated up to some high temperature. Because of the high viscosity, the kinetics of melting is extremely slow in selenium. Therefore, the cell must be heated up very uniformly to avoid small temperature gradients that will crack the cell due to expansion on local melting. We melted selenium by heating the entire bath filled with oil and keeping it at the melting temperature (217 °C) for 5 to 6 hours. The melting process is readily monitored using the acoustical signals. After melting, the temperature was raised by about ten degrees centigrade, and then slowly cooled down to supercool liquid selenium as deeply as possible. At this stage the temperature was controlled by the electrical heater, and the thermostat was turned off.

To reach the amorphous state the heater was turned off when the temperature of the liquid selenium was about 225 °C. Simultaneously the pump of the thermostat with room-temperature oil was turned on, and the wool thermal isolation was removed. With this procedure, the sample cooled down to room temperature in less than 2 minutes. For the measurements with amorphous selenium, the sample temperature was controlled by the thermostat.

The pulsed, phase-sensitive technique requires continuous tracking of the phase of the transmitted signal. During the quenching process the phase was usually lost, and it was reestablished via alignment of the leading edges of transmitted and reflected signals (see Ref. 28 for details). Values of the sound velocity were calculated starting from the amorphous state, where the data of Sogo et al. [26] were used for reference, to the liquid state, where an additional correction of the phase was performed using the data of Gitis and Michailov at temperature 220°C [25]. This second correction did not exceed one period of the rf oscillations, as was also the case in our previous experiments with mercury on crossing the coexistence curve [28, 29]. In terms of sound velocity the correction was less than 40 m/s (4%). If the quenched sample were polycrystalline, the correction would have been much larger because the speed of sound in polycrystalline selenium (3350 m/s [30]) is almost two times larger than that in glassy selenium. These results confirm that the quenched sample was in the amorphous state.

The experiments with longitudinal sound were performed at frequencies of 7 and 22 MHz. The data on the longitudinal sound speed $c_l$ are shown in Fig. 1.

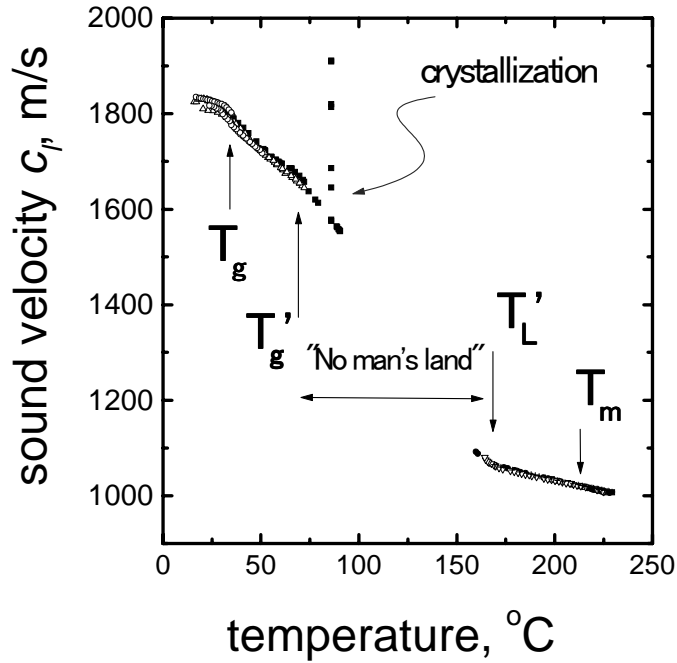

Fig. 1. Speed of longitudinal sound in liquid, supercooled and glassy selenium.
Liquid and supercooled states, experimental data: □, ●, ▽, cooling (7 MHz); ✚, cooling (22 MHz). Glassy state, experimental data: ■, heating (7 MHz); △ , heating and cooling (7 MHz); ○, heating and cooling (22 MHz). $T_g$ is the glass transition temperature, $T_g´$ and $T_L´$ mark stability limits for the superheated glass and supercooled liquid, respectively; $T_m$ is the melting temperature.

The measurements shown in Fig. 1 for the liquid and supercooled states are very reproducible. The reproducibility of the data obtained in the amorphous state is not as good probably because of the non-equilibrium nature of this state.

Four specific temperatures can be defined over the temperature range studied:

(1) The melting temperature, $T_m$. No specific feature in the sound velocity was found at $T_m$, which is consistent with a general definition of a metastable state [31].

(2) The limit of stability of the supercooled liquid state, $T_L´$. At $T_L´ \approx 168$ °C the slope of the temperature dependence abruptly changes. Below this point stationary measurements are no longer possible. Selenium crystallizes practically instantly. It is interesting that selenium, which has viscosity at the melting temperature that is 500 times greater than the viscosity of sulfur, can be supercooled over about the same temperature interval (40 °C for sulfur [9] and 50 °C for selenium). In addition, the crystallization in both these substances progresses at about the same rate. This result probably means that the viscosities of supercooled selenium and sulfur near $T_L'$ are of the same order of magnitude. If the viscosities are similar, then one may speculate that supercooled selenium undergoes a depolymerization transition under cooling, similar to that which occurs in liquid sulfur. Such a transition has been discussed theoretically by Eisenberg and Tobolsky [32].

(3) The glass transition temperature, $T_g$. A step-like change in the sound velocity was found near this point. This feature is discussed below.

(4) The limit of stability of the superheated glassy state, $T_g'$. At temperatures above $T_g'$ ≈ 70 °C the superheated glassy selenium crystallizes. In contrast to the crystallization on cooling near $T_L'$, crystallization above $T_g'$ progresses much more slowly. For this reason, it was possible to take data during this process. Some data are shown in Fig. 1 where time interval between neighboring points is 5 to10 minutes. The rate of crystallization decreases with time, so it could take hours or days (depending on the proximity to $T_g'$) to complete the process. The temperature dependence of the sound velocity exhibits another change in slope near $T_g'$; however, this change is significantly less pronounced than that at $T_L'$.

The temperature dependence of $c_l$ near the glass transition temperature is shown on an expanded scale in Fig. 2. Under heating, the velocity $c_l$ exhibits a step-like drop of about 2 %. This anomaly is centered at 35 °C in agreement with the value of the glass transition temperature measured in the calorimetric, viscometer, and neutron diffraction experiments [18-23]. The sample was heated to 70 °C (open triangles) and to 50 °C (open circles) and then cooled down at the same rate of about 0.5 K/min. The anomalous feature is significantly less pronounced under cooling, and it depends on the maximum heating temperature. Similar hysteresis is often observed in measurements near the glass transition temperature and attributed to the non-equilibrium nature of the amorphous state [2]. The fact that the data obtained under cooling depend on the heating temperature suggests that the sample is partially annealed. For this reason, the data obtained under cooling are difficult to interpret and they are not used in further discussion.

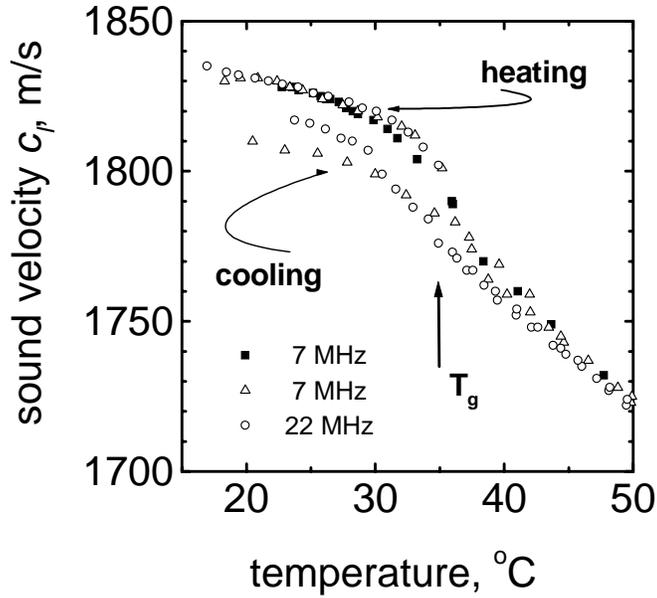

Fig. 2. Longitudinal sound velocity in amorphous selenium near the glass transition temperature.

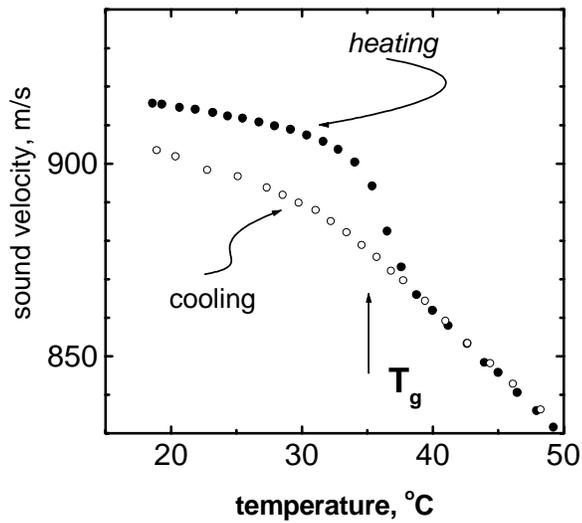

Fig. 3 Velocity of transverse sound in amorphous selenium near the glass transition temperature.

Data for the velocity of the transverse sound $c_t$ in amorphous selenium are shown in Fig. 3. The sample was heated up to 61 °C and cooled down with the same rate of 0.5 K/min. The frequency of the sound was 7 MHz. The observed anomaly near the glass transition temperature has the same character as for the longitudinal sound, although its magnitude (3%) is greater.

## 3. DISCUSSION

Fig. 1 can be interpreted as a phase diagram for supercooled selenium. In the temperature interval from 70 to 168 °C (the so called "no man's land" [6]) it is impossible to maintain a disordered state of selenium that is metastable over long times. This instability interval has a sharp upper boundary at $T_L'$, which can be determined to within 1 K. The curve $c_l$(T) has a pronounced change in the slope at this point. The lower boundary at $T_g'$ less well defined, and the feature in $c_l$(T) is much weaker. Undoubtedly this difference in behavior is associated with the big difference in viscosity at $T_g'$ and $T_L'$. The upper limit of instability for supercooled selenium reported in Ref. 23 (440 K) agrees very well with the value of $T_L'$ found in this work, but the temperature of the lower boundary differs from our estimate by 17 K. This difference is probably due to the lack of a precise experimental definition of $T_g'$ and its dependence on heating and cooling rates.

From the longitudinal and transverse velocities of sound, one can calculate the bulk (*K*), shear (*μ*), and Young (*E*) moduli, and the Poisson ratio (*σ*) using the well-known formulas:

$$K = \rho(c_l^2 - \frac{4}{3}c_t^2),\qquad(1)$$

$$\mu = \rho \cdot c_t^2,\qquad(2)$$

$$E = \frac{9K\mu}{3K + \mu},\qquad(3)$$

$$\sigma = \frac{1}{2} \cdot \frac{3K - 2\mu}{3K + \mu},\qquad(4)$$

where $\rho$ is the mass density.

The density of amorphous selenium at 25°C is 4.3 g/cm$^3$ [26]. Because the precision of the available data for the temperature dependence of the density in selenium [33] is not high enough, it is safer to calculate the ratios of the moduli to the density. The data obtained under increasing temperature were used for these calculations.

Normalized values of the bulk and shear moduli divided by the density are shown in Fig. 4. The values of $K/\rho$ and $\mu/\rho$ at 20 °C are 22 kbar/(g/cm$^3$) and 8.4 kbar/(g/cm$^3$), respectively. Near the glass transition temperature radical changes should occur [34] in both the bulk and shear moduli. As shown in Fig.4, the magnitude of the changes is different for the two quantities. The decrease in the shear modulus is more than twice as large as that in the bulk modulus.

The values of the Young modulus divided by the density and the Poisson ratio are shown in Fig. 5. As expected, both of these quantities experience dramatic changes near $T_g$. Between 30 and 40 centigrade the Young modulus decreases by about 10 %, and the Poisson ratio increases by about 2 %. All three moduli and the Poisson ratio are essentially independent of temperature below $T_g$.

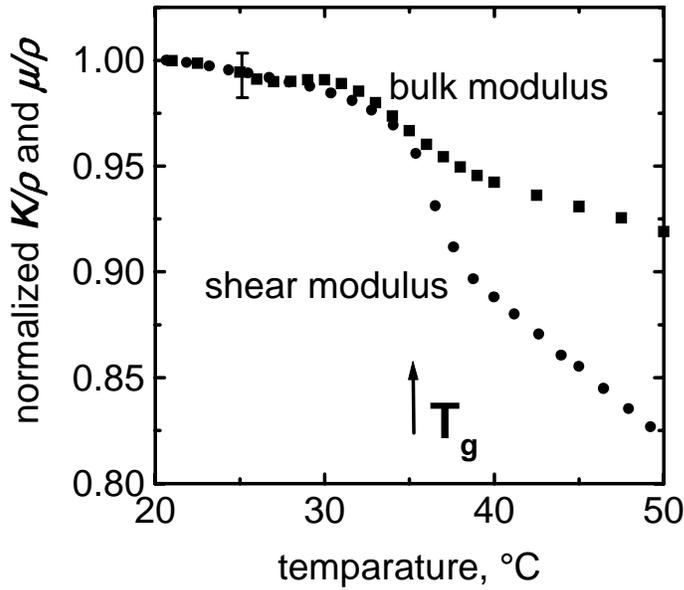

Fig. 4. Normalized values of the bulk and shear moduli divided by the density near the glass transition temperature. The error bar denotes the mean standard deviation resulting from a polynomial fit to the experimental data.

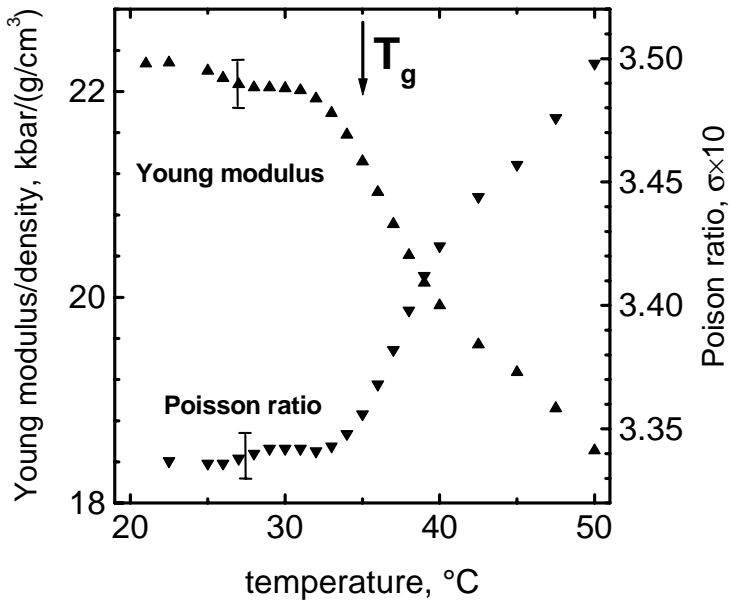

Fig. 5. The Young modulus normalized to the density and the Poison ratio in amorphous selenium near the glass transition. As in Fig. 4, the error bars denote the mean standard deviation resulting from a polynomial fit.

SUMMARY

The temperature dependence of the velocity of sound in supercooled selenium can be interpreted as a phase diagram. The phase diagram contains a melting temperature, at which no feature occurs in the sound velocity on heating from, or cooling to, the supercooled state. In contrast, a step-like anomaly occurs in both the longitudinal and transverse velocities of sound near the glass transition temperature. We speculate that this anomaly in the sound velocity may be a general feature of the glass transition since a similar anomaly has been found in other systems. This speculation should be checked by performing experiments on other glass-forming systems.

We have identified two temperatures, $T_g'$ and $T_L'$, that define a temperature range within which disordered selenium is unstable against crystallization on laboratory time scales. At the upper temperature, $T_L'$, which is well defined, there is a dramatic change in the temperature derivative of the sound velocity. At the lower limit, $T_g'$, which is not well defined, there is only a small change in the temperature derivative of the sound velocity. The nearly instantaneous crystallization that occurs below $T_L'$ suggests that a "depolymerization" transition might occur in supercooled selenium.

ACKNOWLEDGEMENT

The authors acknowledge the donors of the Petroleum Research Fund, administrated by the American Chemical Society (under grant # 326802-AC5), for support of this research. PCT acknowledges support from the National Science Foundation under grant number DMR-0073004. VFK gratefully acknowledges fellowship of the Katholike Universiteit Leuven.